\documentclass[a4paper,11pt]{article}
\usepackage[utf8]{inputenc}
\usepackage{graphicx}
\usepackage{hyperref}
\usepackage{latexsym}
\usepackage{color}
\usepackage{amsmath}
\usepackage{geometry}
\usepackage{slashed}

\def\msbar{{\overline{\rm MS}}}
\newcommand{\abs}[1]{\lvert #1\rvert}

\title{Charmed and $\phi$ meson decay constants from 2+1-flavor lattice QCD}
\author{Ying Chen$^{1,2}\thanks{cheny@ihep.ac.cn}$, Wei-Feng Chiu$^1$, Ming Gong$^{1,2}$, Zhaofeng Liu$^{1,2}$\thanks{liuzf@ihep.ac.cn}, Yunheng Ma$^{1,2}$\\
($\chi$QCD Collaboration)}
\date{}

\begin{document}

\maketitle
\begin{center}
$^1$Institute of High Energy Physics, Chinese Academy of Sciences, Beijing 100049, China\\
$^2$School of Physics, University of Chinese Academy of Sciences, Beijing 100049, China
\end{center}

\begin{abstract}
On a lattice with 2+1-flavor dynamical domain-wall fermions at the physical pion mass, we calculate the decay constants
of $D_{s}^{(*)}$, $D^{(*)}$ and $\phi$.
The lattice size is $48^3\times96$,
which corresponds to a spatial extension of $\sim5.5$ fm with the lattice spacing $a\approx 0.114$ fm. For the valence light, strange and charm quarks, we use overlap fermions at several mass points
close to their physical values.
Our results at the physical point
are $f_D=213(5)$ MeV, $f_{D_s}=249(7)$ MeV, $f_{D^*}=234(6)$ MeV, $f_{D_s^*}=274(7)$ MeV,
and $f_\phi=241(9)$ MeV.
The couplings of $D^*$ and $D_s^*$ to the tensor current ($f_V^T$) can be derived, respectively, from the ratios $f_{D^*}^T/f_{D^*}=0.91(4)$ and $f_{D_s^*}^T/f_{D_s^*}=0.92(4)$, which are the first lattice
QCD results.
We also obtain the ratios $f_{D^*}/f_D=1.10(3)$ and $f_{D_s^*}/f_{D_s}=1.10(4)$, which reflect
the size of heavy quark symmetry breaking in charmed mesons. The ratios
$f_{D_s}/f_{D}=1.16(3)$ and $f_{D_s^*}/f_{D^*}=1.17(3)$ can be taken as a measure of SU(3) flavor symmetry breaking.
\end{abstract}

\section{Introduction}
Meson decay constants are important nonperturbative quantities for the study of meson leptonic decays, and their results from lattice
Quantum Chromodynamics (QCD) have received much attention. The pseudoscalar meson decay constants ($f_P$) can be neatly used to
determine the Cabibbo-Kobayashi-Maskawa (CKM) matrix elements, if combined with experiment measurements of the corresponding leptonic decays.
The newest lattice QCD average of $f_P$ can be found in the review by Flavor Lattice Averaging Group (FLAG)~\cite{Aoki:2019cca}

In principle vector meson decay constants $f_V$ can also be used to determine CKM matrix elements although experimental measurements of leptonic decays of vector mesons
are much harder than those of pseudoscalar mesons due to small branching ratios.
With increasing statistics the leptonic decay of $D_s^*$ may be expected to be measured by BES-III
or Belle II in the near future for the first time for a vector meson~\cite{Donald:2013sra}.
Then the comparison of $f_{D_s^*}$ from experiment and theoretical calculation can be used to study the low energy properties of QCD.

Furthermore, decay constants of heavy-light vector mesons can be used to
test the accuracy of heavy quark effective theory (HQET).
Neglecting terms of $\mathcal{O}(1/m_Q)$, where $m_Q$ is the heavy quark mass,
one has $f_V/f_P=1-2\alpha_s(m_Q)/(3\pi)$~\cite{Neubert:1993mb} from the leading order QCD calculation, which implies that the ratio $f_V/f_P$ approaches one
since the strong coupling constant $\alpha_s(m_Q)$ vanishes in the infinite
heavy quark mass limit. We can obtain the corrections from the higher order terms
in charmed mesons through the ratio $f_V/f_P$ from lattice QCD calculations.
Also, the ratios $f_V/f_P$ for charmed mesons are input parameters for QCD factorization studies of charmed nonleptonic $B$ meson decays~\cite{Keum:2003js,Zou:2009zza}.
Another important quantity $f_V^T$ is the coupling of a vector meson to the tensor current. The nonperturbative determination of the ratio $f_V^T/f_V$ is important in light cone QCD sum rule (LCSR) calculations of form factors in $B$ to vector meson
semileptonic decays (see discussions in \cite{Becirevic:2003pn,Ali:1993vd,Ball:1997rj}).

In this paper, we present a lattice calculation of
$D_{s}^{(*)}$, $D^{(*)}$ and $\phi$
meson decay constants in a lattice setup with chiral fermions, which are
usually expected to be important when light flavors are involved since chiral symmetry is a fundamental property of QCD.
We use overlap fermions for valence quarks and carry out the calculation on 2+1-flavor domain wall fermion gauge configurations generated by the RBC-UKQCD Collaborations. The lattice size is big enough ($\sim5.5$ fm) to avoid large finite volume effects. The light sea quark mass is almost at the physical point.
There have been four lattice QCD calculations of $f_{D_s^*}$ in literatures so far.
Two of them were performed on $2$-flavor gauge
ensembles~\cite{Becirevic:2012ti,Blossier:2018jol}. The other two were
performed on 2+1-flavor ensembles~\cite{Donald:2013sra} and 2+1+1-flavor ensembles~\cite{Lubicz:2017asp}, respectively.
An unexpected large quenching effect of the strange quark was observed in
$f_{D_s^*}$ and $f_{D_s^*}/f_{D_s}$ from the 2-flavor result~\cite{Becirevic:2012ti}
(confirmed in~\cite{Gambino:2019vuo} but with a reduced effect).
While the 2-flavor result from~\cite{Blossier:2018jol} shows a much less pronounced effect.
In this study we give an independent 2+1-flavor calculation for $f_{D_s^*}$ to compare with the aforementioned calculations.

The rest of this paper is organized as follows. In Sec.~\ref{sec:framework} we give our framework of the calculation, including the definitions
of the decay constants and the lattice setup. Sec.~\ref{sec:details} presents
the details of the analyses, the numerical results and discussions. Finally, we summarize in Sec.~\ref{sec:summary}.

\section{Definitions and lattice setup}
\label{sec:framework}
\subsection{Decay constants of pseudoscalar and vector mesons}
The decay constant $f_{P}$ of a pseudoscalar meson $P$ is defined through
\begin{equation}\label{fps}
\langle 0 | \bar \psi_1(x) \gamma_\mu \gamma_5 \psi_2(x) |P(p) \rangle =ip_\mu f_{P} e^{-ipx},
\end{equation}
with $p_\mu $ being the momentum of the meson. By using the partially conserved axial vector current (PCAC) relation,
$f_{P}$ can be obtained from the matrix element of the pseudoscalar density
\begin{equation}
(m_{1}+m_{2})\langle 0 | \bar \psi_1(0) \gamma_5 \psi_2(0) |P(p) \rangle = m_{P}^2 f_{P},
\label{eq:fp_pseudo}
\end{equation}
where $m_{1,2}$ are quark masses and $m_P$ is the pseudoscalar meson mass.
For overlap fermions, the quark mass and pseudoscalar density $ \bar \psi_1 \gamma_5 \psi_2 $ renormalization constants
cancel each other ($Z_P=Z_m^{-1}$) due to chiral symmetry. This makes $f_P$ obtained from Eq.(\ref{eq:fp_pseudo}) free of renormalization.

The vector meson decay constant $f_V$ is given by the matrix element of the vector current between the vacuum and vector meson $V$ as
\begin{equation}
\langle 0|\bar\psi_1(0)\gamma_{\mu}\psi_2(0)|V(p,\lambda)\rangle = m_{V}f_V \epsilon_{\mu}(p,\lambda),
\label{eq:fv}
\end{equation}
where $\epsilon_{\mu}(p,\lambda)$ is the polarization vector of meson $V(p,\lambda)$ with helicity $\lambda$.
We use the local vector current on the lattice to compute the above matrix element for convenience.
The price to pay is the need of a calculation of the finite renormalization constant for the local current, which was obtained nonperturbatively
in Ref.~\cite{Bi:2017ybi} for our lattice setup.

Besides $f_V$, vector mesons have another decay constant $f_V^T$ which is defined through the following matrix element of the tensor current
\begin{equation}
 \langle 0|\bar\psi_1(0)\sigma_{\mu\nu}\psi_2(0)|V(p,\lambda)\rangle = i f_V^T (\epsilon_{\mu}(p,\lambda) p_\nu - \epsilon_{\nu}(p,\lambda) p_\mu).
\label{eq:fvt}
\end{equation}
Here in the tensor current $\sigma_{\mu\nu}=(i/2)[\gamma_\mu,\gamma_\nu]$. Since the tensor current has a nonzero anomalous dimension,
we will give values of $f_V^T$ in the commonly used $\msbar$ scheme and at a scale $\mu=2$ GeV. The matching factor from the lattice
to the continuum $\msbar$ scheme for the tensor current was presented in Ref.~\cite{Bi:2017ybi}.

\subsection{Lattice setup}
Our calculation is carried out on the gauge configurations of $N_f = 2+1 $ domain wall fermions generated by the RBC-UKQCD Collaborations~\cite{Blum:2014tka}.
We use the gauge ensemble named as 48I with lattice size $L^3 \times T = 48^3 \times 96 $ and
pion mass $m^{\rm (sea)}_{\pi} = 139.2(4)$ MeV
from the sea quarks.
The lattice spacing was determined to be $ a^{-1}=1.730(4) \mbox{ GeV}$~\cite{Blum:2014tka}, thus the spatial extension of the lattice
is about $La \sim 5.5 \mbox{ fm}$.
The parameters of the configurations are given in Table~\ref{tab:param}.
\begin{table}
\begin{center}
\caption{Parameters of gauge configurations used in this work. $am_q^{\rm (val)} (q=l,s,c)$ are
 the valence quark mass parameters in lattice units and
the corresponding pion masses (in MeV) are from Ref.~\cite{Sun:2018cdr}.
The physical charm quark mass $am_c^{\rm phy}$ is estimated to be around 0.73 (see below).}
\begin{tabular}{ll}
\hline\hline
   $L^3\times T$  & $48^3\times96$ \\
   $a^{-1}$(GeV) &  1.730(4) \\
   $N_{\rm conf}$ & 45 \\
\hline
   $am^{\rm (val)}_l$ & 0.0017, 0.0024, 0.0030, 0.0060 \\
   $m_\pi/$MeV & 114(2), 135(2), 149(2), 208(2) \\
   $am^{\rm (val)}_s$ &   0.0580, 0.0650 \\
   $am^{\rm (val)}_c$  & 0.6800, 0.7000, 0.7200, 0.7400   \\
\hline\hline
\end{tabular}
\label{tab:param}
\end{center}
\end{table}

We use overlap fermions for valence quarks to perform a mixed action study.
The mismatch of the mixed valence and sea pion masses between the domain-wall fermion and the overlap fermion,
measured by $\Delta_{\rm mix}$, is $0.030(6)(5)$ GeV$^4$~\cite{Lujan:2012wg}, which is very small reflecting a small partial quenching effect.
The multi-mass algorithm of overlap fermions~\cite{Alexandru:2011sc} permits calculations of multiple quark propagators with
a reasonable cost. We calculate propagators with a range of masses from the light to charm quark on 45 configurations. The valence quark masses $am_q^{\rm (val)} (q=l,s,c)$ in lattice units are given in Table~\ref{tab:param}.
The deflation algorithm is adopted to accelerate the inversion by projecting out the 1000 low eigenvectors (including zero modes) of the overlap Dirac operator, which are calculated explicitly
beforehand.

We use four mass parameters $am_l^{\rm (val)}$ (as listed in Table~\ref{tab:param}) for the light valence quarks for chiral interpolation. The corresponding pion masses range from $114$ MeV to $208$ MeV~\cite{Sun:2018cdr}.
Two strange quark mass parameters are used to extrapolate to the physical strange quark mass point.
The bare charm quark masses that we use are around $0.72$ in lattice units, which are not small.
Although for chiral lattice fermions the discretization error due to the heavy quark mass starts at $\mathcal{O}((am_c)^2)$,
it could still be large. Thus, we shall try to estimate the finite lattice spacing effects in our results for $D$-mesons.

\subsection{Two-point correlators}
The matrix elements in Eq.~(\ref{fps}),~(\ref{eq:fv}) and ~(\ref{eq:fvt}), from which the decay constants are defined, can be derived directly from the related two-point functions with the currents being the sink operators. Since the mesons involved in this study are all the ground state hadrons, in order for the matrix elements to be determined precisely, it is desired that the two-point functions are dominated by the contribution from the ground states. In this work, we adopt the Coulomb wall-source technique. That is to say, we perform the Coulomb gauge fixing to the gauge configurations firstly, and then calculate the two-point functions using the following wall-source operators which are obviously gauge dependent,
\begin{equation}
 O^{(W)}_{\Gamma}(t)= \sum\limits_{\vec{y},\vec{z}}^{}\bar \psi ^{f_1}(\vec{y},t) \Gamma \psi ^{f_2}(\vec{z},t),
\end{equation}
where $\psi ^{f} = u,d,s,...$ and $\Gamma = \gamma_5$ for pseudoscalar mesons and $\Gamma =\gamma _i$ ($i=1,2,3$) for vector mesons.
From our experience~\cite{Sun:2018cdr} besides the suppression of excited states, the choice of wall source can also suppress the $P$-wave scattering states in the vector channels.

For the sink operators, we use spatially extended operators $O_{\Gamma}(\vec{x},t;\vec{r})$ by splitting the quark and anti-quark field
with spatial displacement $\vec{r}$, namely, $O_{\Gamma}(\vec{x},t;\vec{r}) \equiv \bar \psi^{f_1}(\vec{x}+\vec r,t) \Gamma \psi^{f_2}(\vec{x},t)$.
The operators with the same spatial separation $r\equiv\lvert \vec{r} \rvert$ are averaged to guarantee the correct quantum number, and also to increase the statistics as a by-product. Thus, the two-point functions we calculate are
\begin{equation}
C_{P}(r,t)= \frac{1}{N_r}\sum\limits_{\vec{x},\abs{\vec{r}} = r} \langle 0 | O_{\gamma_5}(\vec{x},t;\vec{r})O^{(W)\dagger}_{\gamma_5}(0)|0\rangle,
\end{equation}
\begin{equation}
C_{V}(r,t)= \frac{1}{3 N_r}\sum\limits_{\vec{x},i,\abs{\vec{r}} = r} \langle 0 | O_{\gamma_i}(\vec{x},t;\vec{r})O^{(W)\dagger}_{\gamma_i}(0)|0\rangle,
\end{equation}
and
\begin{equation}
 C_T(r=0,t)=\frac{1}{3}\sum_{\vec x,i}\langle 0 | O_{\sigma_{0i}}(\vec x,t)O^{(W)\dagger}_{\gamma_i}(0)|0\rangle,
\end{equation}
where $N_r$ is the number of $O_\Gamma(\vec x,t;\vec{r})$'s with the same $\abs{\vec{r}}= r$. The two-point functions
$C(r,t)$ with different $r$ can be calculated simultaneously without expensive extra inversions.
After the insertion of the intermediate states,
the spectral expression of a two-point function reads
\begin{equation}\label{spec-decomp}
C(r,t) = \sum\limits_{n,|\vec{r}|=r} \frac{1}{2m_n N_r}\langle 0|O_\Gamma(\vec{0},0;\vec{r})|n\rangle \langle n|O^{(W)\dagger} |0\rangle e^{-m_n t}\equiv \sum\limits_n \Phi_n(r)e^{-m_n t},
\end{equation}
where $\Phi_n (r)$ is proportional to the Bethe-Salpeter amplitude $\frac{1}{N_r}\sum\limits_{|\vec{r}|=r}\langle 0|O_\Gamma(\vec{0},0;\vec{r})|n\rangle$ for the $n$-th state. Since the $r$ dependences of $\Phi_n(r)$ are different for different states in each channel, a proper linear combination of several $C(r,t)$'s with different $r$ may give an optimal two-point function $C(\omega,t)\equiv\sum\limits_{\omega_i} \omega_i C(r_i,t)$ which
is dominated by the ground state.

Obviously, the parameterization of Eq.~(\ref{spec-decomp}) shows that the spectral weight $\Phi_n(r=0)$ is proportional to the matrix element that defines the decay constant of a specific meson state. However, in order to get the decay constant, we need to remove the factor  $\langle n | O^{(W)\dagger}_{\Gamma}|0\rangle$, which is the matrix element of the wall-source operator
 $O^{(W)\dagger}$ between the vacuum and the meson state and can be derived from the wall-to-wall correlation function
\begin{equation}
C^{W}(t) =  \langle 0 | O^{(W)}(t)O^{(W)\dagger}(0)|0\rangle.
\end{equation}

\section{Numerical analyses}
\label{sec:details}
\subsection{Meson masses}
To extract the meson masses, we apply two fitting strategies. One strategy is applying
correlated simultaneous fittings to the correlation functions with different $r$'s using one
(for vector mesons) or two (for pseudoscalar mesons) mass terms. The function form used in the simultaneous fits is
\begin{equation}
C(r,t) = \sum\limits_{n=0} \Phi_n(r) \left[ e^{-m_n t} + e^{-m_n(T-t)} \right],
\end{equation}
where $T=96$ and the second term in the brackets on the right hand side comes from the propagation of the correlator in the negative time direction. $\Phi_n(r)$ and $m_n$ are fitted with the minimum $\chi^2$ method.
We vary the number of mass terms to two or three and check the
stability of the fitting results.
Within statistical uncertainties the fitted ground state mass $m_0$
does not depend on the number of mass terms. The upper limit of the
fitting range $[t_{\rm min}, t_{\rm max}]$
is chosen by the following criteria. For the pseudoscalar channel $t_{\rm max}$ is fixed to
the maximum value where the relative errors of correlators
satisfy $\delta C/C\le 5\%$. For the vector channel
$t_{\rm max}$ is chosen by requiring $\delta C/C\le 10\%$. The lower limit of
the fitting range is varied in
a wide range when doing the fittings and we check the stability of the results.
Among all the fittings which have $\chi^2/{\rm dof}\le1.0$ and give a consistent ground state mass we then choose the earliest $t_{\rm min}$
to give our final results.
The uncertainties are obtained from Jackknife analyses to take into account the correlations among the data as we repeat the fitting
for each Jackknife ensemble.

\begin{figure}[htpb]
\centering\includegraphics[width=0.48\textwidth]{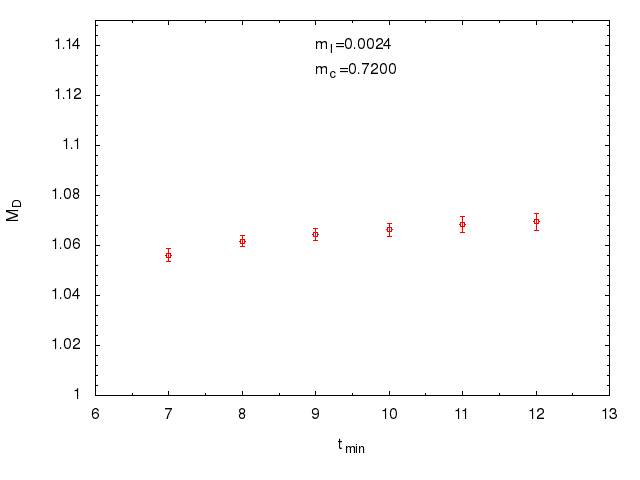}
\centering\includegraphics[width=0.48\textwidth]{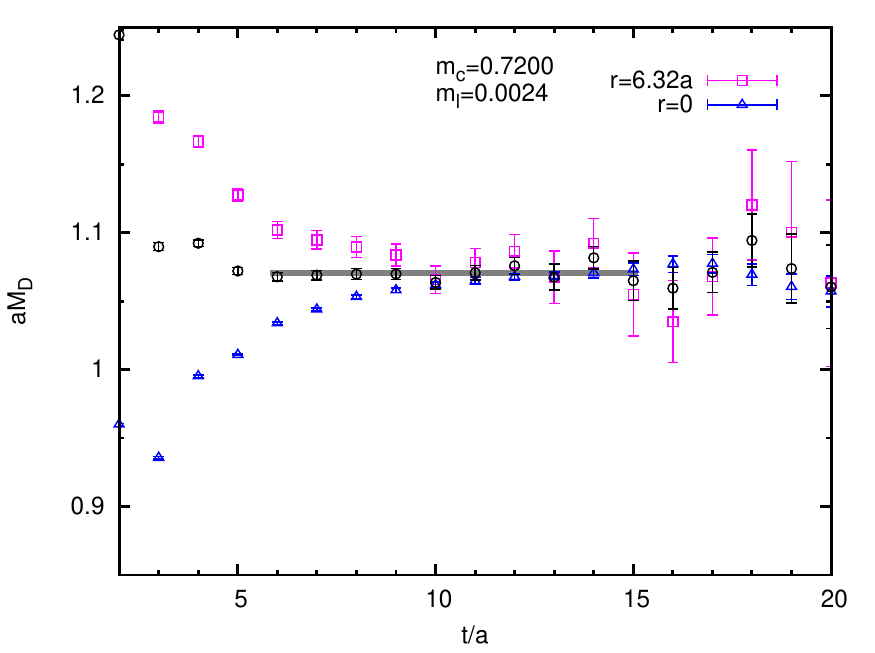}
\caption{$M_{D}$ in lattice units as a function of $t_{\rm min}$ (left panel).
$M_{D}$ (the band in the right graph) from fitting range [11, 18]
is compared with the corresponding
effective masses from varies correlators (right panel).}
\label{fig:MD}
\end{figure}

In the left panel of Fig.~\ref{fig:MD} we show the fitted
ground state mass $M_{D}$ in lattice units as a function of $t_{\rm min}$.
Here we finally choose the fitting range [11, 18] for the $D$ meson.
In the right panel of Fig.~\ref{fig:MD} the obtained ground state mass $M_{D}$ (the band
in the graph) is compared with the corresponding
effective masses $M_{\rm eff}=\log(C(r,t)/C(r,t+1))$ from various correlators
with different $r$. The data points in magenta squares are from the correlator with $r=6.32a$
($\vec r=(2, 6, 0)$ and permutations averaged). The ones in blue triangles are from the local sink
correlator with $r=0$. The ones in black circles are the effective masses from a combination of
two correlators
\begin{equation}
C(\omega, t)=C(r=1, t)+\omega C(r, t),
\end{equation}
where we can tune the parameter $\omega$ and use various $C(r, t)$
to make the effective mass plateau from $C(\omega, t)$ appear
as early as possible. This leads to our second fitting strategy. Different states
with a same quantum number contribute differently to the correlators $C_\Gamma(r, t)$.
And these contributions vary as $r$ varies. Thus, it is possible to find a large $r$
such that the contribution of the lowest excited state to $\omega C(r, t)$ cancels that to $C(r=1, t)$ and $C(\omega ,t)$ is dominated by the ground state.

In the right panel of Fig.~\ref{fig:MD}, the black circles show a mass plateau which
starts much earlier than that from the correlator $C(r=6.32a, t)$ or $C(r=0, t)$.
Therefore, we can fit the combined correlator $C(\omega, t)$ easily with a single
exponential term. We check that this fitting gives stable and consistent ground state mass
as we vary the parameter $\omega$. We also confirm that the results from the above two
fitting strategies are in consistency.

The fitting results of $am_D$ and $am_{D^*}$ from the two strategies with $am_c=0.72$
are summarized in Table~\ref{tab:compare} for comparison.
The second strategy gives smaller statistical
uncertainties since the mass plateau from the combined correlator appears earlier
and thus data points with less errors are used in fittings.
Similar advantages of the second strategy are observed in the analyses of other meson masses,
therefore we adopt strategy II to obtain meson masses in the following.
\begin{table}
\begin{center}
\caption{The masses of $D$-mesons with $am_c=0.72$ extracted from two fitting strategies.
The first errors are statistical from Jackknife analyses. The second errors
are systematic errors from variations in the center values as we vary $t_{\rm min}$.
The two strategies give consistent results.}
\begin{tabular}{cccccl}\hline\hline
  $am_q$       & 0.0017 & 0.0024 & 0.0030 & 0.0060 &    \\
  \hline
  $am_{D}$  & 1.070(4)(1) & 1.070(3)(1) & 1.070(3)(1) & 1.071(3)(1) &   strategy I \\
            & 1.071(2)(1) & 1.071(2)(1) & 1.071(2)(1) & 1.073(1)(1) & strategy II  \\
  \hline
  $am_{D^*}$ & 1.156(8)(1) & 1.157(8)(1) & 1.158(7)(2) & 1.160(6)(2) & strategy I  \\
             & 1.160(2)(1) & 1.160(2)(1) & 1.160(2)(1) & 1.162(2)(1) & strategy II \\
  \hline
  \hline
\end{tabular}
\label{tab:compare}
\end{center}
\end{table}

The results of the pion and kaon masses are shown in Table~\ref{tab:k_pi}.
The pion mass and the combination $m_{ss}^2\equiv 2m_K^2-m_\pi^2$ are used to
fix the physical up (degenerate with the down quark) and strange quark mass respectively.
From Table~\ref{tab:k_pi} we can see that $2m_K^2-m_\pi^2$ is independent of the pion mass
(or equivalently the up/down quark mass) within the statistical uncertainties.
This is exactly what we expect from the lowest-order analysis of chiral perturbation theory
and it is the reason why we use this combination.
\begin{table}
\begin{center}
\caption{Masses of pion and kaon with statistical uncertainties from Jackknife analyses.}
\begin{tabular}{ccccc}\hline\hline
$am_s$ & $am_q$ & $am_K$ &  $am_\pi$ & $a^2(2m_K^2-m_\pi^2)$  \\
\hline
0.0580 & 0.0017  & 0.2608(24) &  0.0659(12) & 0.1317(25)  \\
 & 0.0024  & 0.2621(20)  & 0.0780(12) & 0.1313(21)   \\
 & 0.0030  & 0.2631(19)  & 0.0861(12) & 0.1310(20)  \\
 & 0.0060  & 0.2689(20)  & 0.1202(12) & 0.1302(22)  \\
\hline
0.0650 & 0.0017  & 0.2755(22)  & 0.0659(12) & 0.1475(24)  \\
 & 0.0024 & 0.2769(22)  & 0.0780(12) & 0.1473(24)  \\\
 & 0.0030 & 0.2780(21)  & 0.0861(12) & 0.1472(23)  \\
 & 0.0060 & 0.2833(18)  & 0.1202(12) & 0.1461(21)  \\
  \hline
  \hline
\end{tabular}
\label{tab:k_pi}
\end{center}
\end{table}
The results of the meson masses and decay constants will be interpolated/extrapolated to
the physical point where $(a^2m^2_\pi)^{\rm phys}=0.00651(3)$ and
$a^2m_{ss}^2({\rm phys})\equiv a^2(2m_K^2-m_\pi^2)^{\rm phys}=0.1565(6)$ by using
$m_\pi^{\rm phys}=139.6$ MeV and $m_K^{\rm phys}=493.7$ MeV~\cite{Tanabashi:2018oca}.
Here the uncertainties come from the error of the lattice spacing.
Since these uncertainties are much smaller than our statistical error or
the discretization error as we will see later,
we ignore them in our estimate of the systematic uncertainty.

In Table~\ref{tab:kstar} we collect the masses of $\phi$ and $K^*$ at our valence quark masses.
\begin{table}
\begin{center}
\caption{Masses and decay constants of $\phi$ and $K^*$ with the statistical uncertainties.
The fitting range of correlators for $\phi$ is $t\in [11, 19]$. The range for $K^*$ is $t\in[8, 15]$.}
\begin{tabular}{ccccc}\hline\hline
$am_s$ & $am_\phi$ & $af_\phi^{\rm bare}$ & $am_q$ & $am_{K^*}$ \\
\hline
0.0580 & 0.563(5) & 0.126(7) & 0.0017  & 0.505(8) \\
       &   & & 0.0024  & 0.504(7)  \\
       &  &  & 0.0030  & 0.503(7)  \\
       &   & & 0.0060  & 0.504(6)  \\
\hline
0.0650 & 0.579(5) & 0.127(7) & 0.0017  & 0.514(7)  \\
       &   &  & 0.0024  & 0.512(7)  \\
       &  &  & 0.0030  & 0.511(7)  \\
       &   & & 0.0060  & 0.512(7)  \\
\hline\hline
\end{tabular}
\label{tab:kstar}
\end{center}
\end{table}
From the data we see that the mass of $K^*$ barely depends on the light quark mass
with our current statistical uncertainties.
To obtain $m_{K^*}$ at the physical point, we use the following interpolation/extrapolation form
\begin{equation}
 m_{K^*}(m_\pi,m_{ss})=m_{K^*}^{\rm phys}+b_1\Delta m_\pi^2+b_2\Delta m_{ss}^2,
\label{eq:mKstar}
\end{equation}
where $\Delta m_\pi^2=m_\pi^2-m_\pi^2({\rm phys})$
and $\Delta m_{ss}^2=m_{ss}^2-m_{ss}^2({\rm phys})$.
This is the Taylor expansion around the physical $u/d$ and strange quark masses and we keep
only the lowest order, i.e., the linear terms since our quark masses are close to their
physical values. Then we obtain
\begin{equation}
m_{K^*}^{\rm phys}=895(10)\mbox{ MeV},
\end{equation}
where the error includes the statistical/fit uncertainty and the uncertainty of the lattice spacing.
The parameter $b_1$ from the fitting is consistent with zero within uncertainty as expected from the raw data.

For the mass of $\phi$ we do a linear extrapolation to the physical point
$a^2m_{ss}^2({\rm phys})=0.1565$ since we only have two data points as given
in Table~\ref{tab:kstar}. For the corresponding $a^2m_{ss}^2$ at each of the two strange quark masses
we use the average of the four values in the last column of Table~\ref{tab:k_pi}.
This extrapolation gives
\begin{equation}
m_\phi^{\rm phys}=1.018(17)\mbox{ GeV}
\end{equation}
with lattice spacing error included.
Both $m_{K^*}^{\rm phys}$ and $m_\phi^{\rm phys}$ are in good agreement with their
experiment values. This means that the finite lattice spacing effects in the study of
light hadrons are smaller than our current statistical uncertainties.

Vector mesons can decay to two pseudoscalar mesons through $P$-wave. On our lattice
the minimal nonzero momentum is 226 MeV, which is not small. The thresholds of
$P$-wave decays for $\phi$, $D^*$ and $D_s^*$ mesons are not open on our lattice.
But $K^*$ can decay to $K\pi$ on our lattice. We observed mass plateaus for the $K^*$
meson but not for the scattering states of $K\pi$, which we believe are suppressed by
the usage of Coulomb gauge wall source when calculating the 2-point functions~\cite{Sun:2018cdr}. The agreement of $m_{K^*}^{\rm phys}$ and $m_\phi^{\rm phys}$ (from our
interpolation/extrapolation) with their
experimental values tells us that it is safe to ignore the threshold effects
at our current precision.

\begin{figure}[htpb]
\centering\includegraphics[width=0.43\textwidth]{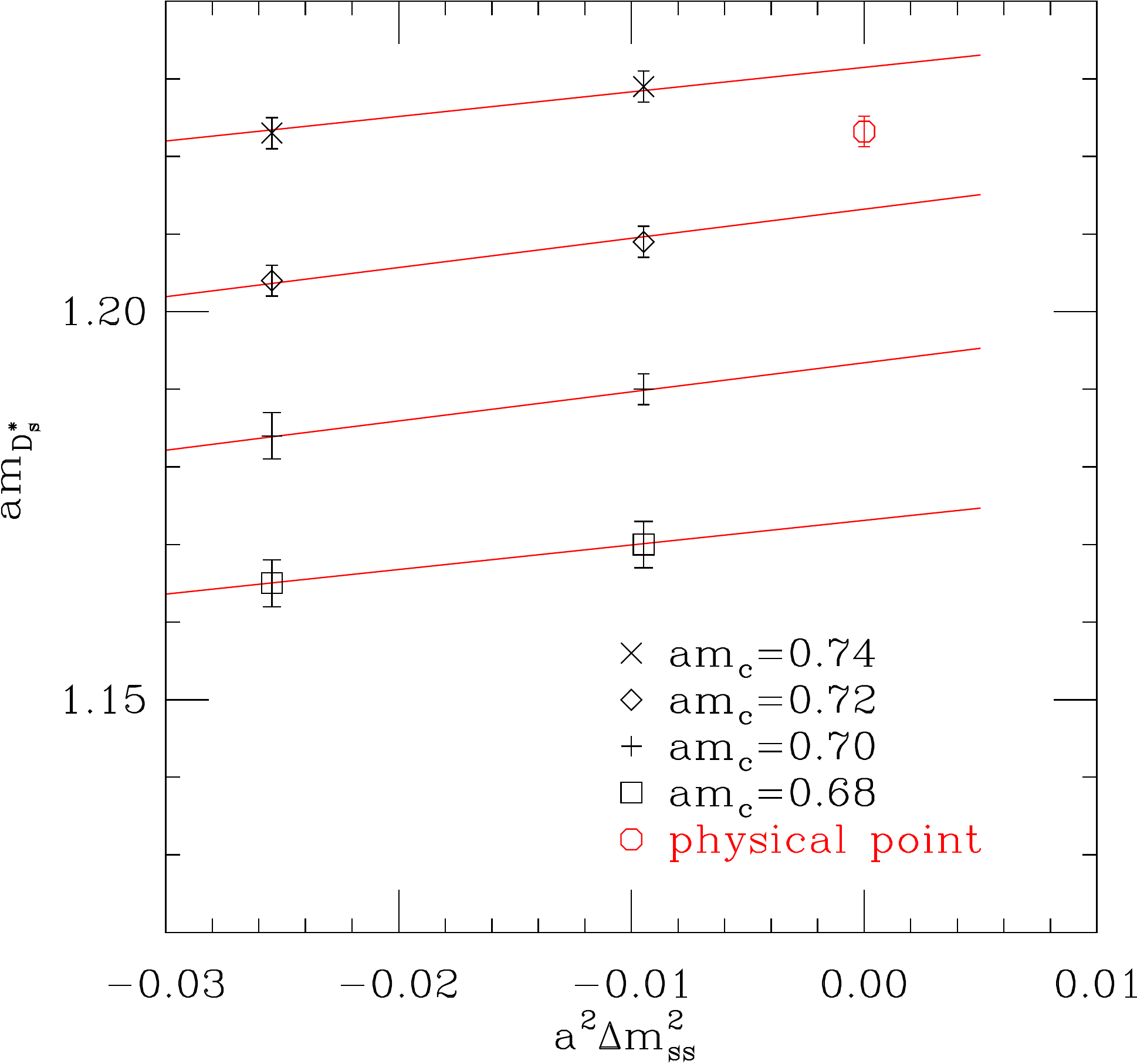}
\centering\includegraphics[width=0.43\textwidth]{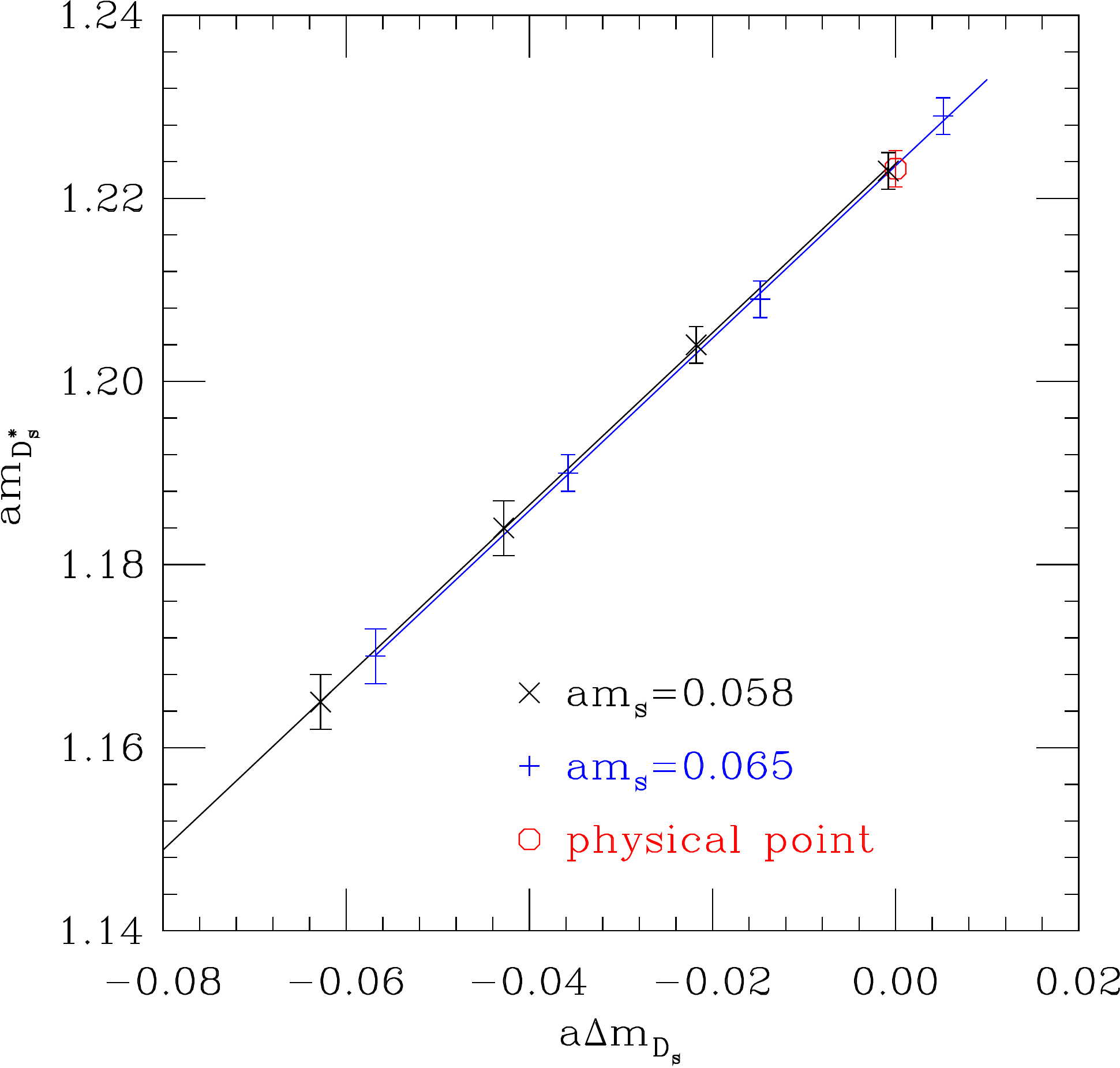}
\caption{The interpolation/extrapolation of $m_{D_s^*}$ to the physical point by using Eq.(\ref{eq:mDsstar}). $am_{D_s^*}$ is plotted
as a function of $a^2\Delta m_{ss}^2$ (left panel) or $a\Delta m_{D_s}$ (right panel).
The octagon is the result at the physical strange and charm quark mass point.}
\label{fig:mDsstar_fit}
\end{figure}

The masses of $D_s$ and $D_s^*$ mesons are listed in Table.~\ref{tab:Ds}.
\begin{table}
\begin{center}
\caption{Masses and decay constants of $D_s$ and $D_s^*$ with statistical uncertainties.
The fitting range of correlators for $D_s$ is $t\in [17, 28]$. The range for $D_s^*$ is $t\in[12, 25]$. The ratio $f_{D_s^*}^{\rm bare}/f_{D_s}$ is collected in the last column.}
\begin{tabular}{ccccccc}\hline\hline
$am_c$ & $am_s$ & $am_{D_s}$ & $af_{D_s}$ &  $am_{D_s^*}$ & $af_{D_s^*}^{\rm bare}$ & $f_{D_s^*}^{\rm bare}/f_{D_s}$  \\
\hline
0.68   & 0.058 & 1.075(1) & 0.139(3)  & 1.165(3) &  0.141(3) & 1.011(27) \\
       & 0.065 & 1.081(1) & 0.141(3)  & 1.170(3) &  0.143(3) & 1.008(25) \\
\hline
0.70 & 0.058 & 1.095(1) & 0.140(3) &  1.184(3) &  0.141(3) & 1.009(27) \\
     & 0.065 & 1.102(1) & 0.142(3) &  1.190(2) &  0.143(3) & 1.005(26) \\
\hline
0.72 & 0.058 & 1.116(1) & 0.140(3) &  1.204(2) &  0.141(3) & 1.007(27) \\
     & 0.065 & 1.123(1) & 0.142(3) &  1.209(2) &  0.143(3) & 1.002(26) \\
\hline
0.74 & 0.058 & 1.137(1) & 0.141(3) &  1.223(2) &  0.141(3) & 1.004(28) \\
     & 0.065 & 1.143(1) & 0.143(3) &  1.229(2) &  0.143(3) & 1.000(27) \\
\hline\hline
\end{tabular}
\label{tab:Ds}
\end{center}
\end{table}
We use the experimental value of $D_s$ (together with $m_{ss}^2({\rm phys})$ in the above)
to set the physical charm (and strange) quark mass. With our lattice
spacing we have $(am_{D_s})^{\rm phys}=1.1378(26)$ by using $m_{D_s}=1968.34(7)$ MeV from Particle Data Group (PDG2018)~\cite{Tanabashi:2018oca}. We use the following function similar to Eq.(\ref{eq:mKstar})
to interpolate/extrapolate $m_{D_s^*}$ to the physical strange and charm quark mass point:
\begin{equation}
 m_{D_s^*}(m_{ss},m_{D_s})=m_{D_s^*}^{\rm phys}+b_2\Delta m_{ss}^2+b_3\Delta m_{D_s},
\label{eq:mDsstar}
\end{equation}
where $\Delta m_{D_s}=m_{D_s}-(m_{D_s})^{\rm phys}$ and $b_3$ is another free parameter.
From this we obtain
\begin{equation}
m_{D_s^*}^{\rm phys}=2.116(6)\mbox{ GeV},
\end{equation}
which agrees with the experiment value $2.1122(4)$ GeV~\cite{Tanabashi:2018oca}. The interpolation/extrapo\-lation
is shown in Fig.~\ref{fig:mDsstar_fit}. The function Eq.(\ref{eq:mDsstar}) can describe the data very well.
The dependence of $m_{D_s^*}$ on the strange quark mass is relatively small. Therefore, the slope of the straight lines in the left plot of Fig.~\ref{fig:mDsstar_fit} is small. This is also the reason why the two lines in the right plot are very close to each other. The dependence on the charm quark mass is apparent. From the position of the physical point in the left plot we can read the physical charm quark mass is around $am_c=0.73$.

The masses of $D$ and $D^*$ mesons are listed in Table.~\ref{tab:D}.
\begin{table}
\begin{center}
\caption{Masses and decay constants of $D$ and $D^*$ with statistical uncertainties.
The fitting range of correlators for $D$ is $t\in [11, 18]$. The range for $D^*$ is $t\in[10, 16]$. The ratio $f_{D^*}^{\rm bare}/f_{D}$ is collected in the last column.}
\begin{tabular}{ccccccc}\hline\hline
$am_c$ & $am_l$ & $am_D$ & $af_D$ &  $am_{D^*}$ & $af_{D^*}^{\rm bare}$ & $f_{D^*}^{\rm bare}/f_{D}$  \\
\hline
0.68   & 0.0017 & 1.028(2) & 0.122(2)  & 1.120(3)  & 0.123(5) & 1.01(4) \\
       & 0.0024 & 1.029(2) & 0.122(2)  & 1.121(2)  & 0.123(4) & 1.01(4) \\
       & 0.0030 & 1.029(2) & 0.122(2)  & 1.121(2)  & 0.123(4) & 1.01(4) \\
       & 0.0060 & 1.030(2) & 0.123(2)  & 1.123(2)  & 0.124(3) & 1.01(3) \\
\hline
0.70 & 0.0017 & 1.049(2) & 0.123(2) & 1.140(3)  &  0.123(5) & 1.00(4) \\
     & 0.0024 & 1.050(2) & 0.123(2) & 1.141(2)  &  0.123(4) & 1.00(4) \\
     & 0.0030 & 1.050(2) & 0.123(2) & 1.141(2)  &  0.123(4) & 1.00(4) \\
     & 0.0060 & 1.052(1) & 0.123(2) & 1.142(2)  &  0.124(3) & 1.01(3) \\
\hline
0.72 & 0.0017 & 1.071(2) & 0.123(2) & 1.160(2)  &  0.123(5) & 1.00(4) \\
     & 0.0024 & 1.071(2) & 0.123(2) & 1.160(2)  &  0.123(4) & 1.00(4) \\
     & 0.0030 & 1.071(2) & 0.123(2) & 1.161(2)  &  0.123(4) & 1.00(4) \\
     & 0.0060 & 1.073(1) & 0.123(2) & 1.162(2)  &  0.123(3) & 1.00(3) \\
\hline
0.74 & 0.0017 & 1.092(2) & 0.123(2) & 1.180(2)  &  0.123(5) & 1.00(4) \\
     & 0.0024 & 1.092(2) & 0.123(2) & 1.180(2)  &  0.123(4) & 1.00(4) \\
     & 0.0030 & 1.092(2) & 0.123(2) & 1.180(2)  &  0.123(4) & 1.00(4) \\
     & 0.0060 & 1.094(1) & 0.124(2) & 1.182(2)  &  0.123(3) & 0.99(3) \\
\hline\hline
\end{tabular}
\label{tab:D}
\end{center}
\end{table}
The following ansatz is used to interpolate/extrapolate our numerical results to
the physical quark mass point:
\begin{equation}
 m_{D^{(*)}}(m_\pi,m_{D_s})=m_{D^{(*)}}^{\rm phys}+b_1\Delta m_\pi^2
 +b_2\Delta m_{ss}^2+b_3\Delta m_{D_s}.
\label{eq:D}
\end{equation}
Here the term $b_2\Delta m_{ss}^2$ appears because our lattice results $m_{D_s}$
are not calculated at the physical strange quark mass and $m_{D_s}^{\rm phys}$ is used
to set the physical charm quark mass. We get
\begin{equation}
m_D^{\rm phys}=1.873(5)\mbox{ GeV}\quad \mbox{and}\quad
m_{D^*}^{\rm phys}=2.026(5)\mbox{ GeV}
\end{equation}
for the two mesons respectively after the interpolations/extrapolations.
Our $D$ meson mass agrees with the PDG2018 value $m_{D^\pm}=1.86965(5)$ GeV within $1\sigma$.
However our $D^*$ meson mass is heavier than the PDG2018 value $m_{D^{*\pm}}=2.01026(5)$ GeV
by about $1\%$. Thus, we estimate the discretization error associated with the large charm quark mass
to be about $1\%$ in our results for the charmed meson masses.

\subsection{Decay constants}
\subsubsection{Renormalization constants}
\label{sec:zavt}
Before we go into the data analyses for the meson decay constants, we present first the
renormalization constants (RCs) for the local vector current and the tensor current.
The RCs of quark bilinear operators for our lattice setup (overlap fermions on domain-wall fermion configurations)
were calculated nonperturbatively in Refs.~\cite{Bi:2017ybi,Liu:2013yxz}. For the 48I ensemble
used in this work we employed
both the RI/MOM and the RI/SMOM schemes to calculate those constants
nonperturbatively~\cite{Bi:2017ybi}.
The matching factors to the $\msbar$ scheme for the local axial vector current
$Z_A$ and for
the tensor current (at scale $2$ GeV) are listed in Table.~\ref{tab:zfactors}.
\begin{table}
\begin{center}
\caption{Matching factors to the $\msbar$ scheme for the local axial vector
current and for the tensor current~\cite{Bi:2017ybi}.}
\begin{tabular}{ccc}
\hline\hline
 $Z_A(=Z_V)$     & $Z_T/Z_A(2\mbox{ GeV})$ & $Z_T(2\mbox{ GeV})$  \\
 1.1025(16) &   1.055(31)        &   1.163(34)         \\
\hline\hline
\end{tabular}
\label{tab:zfactors}
\end{center}
\end{table}
Because we use chiral fermions, we have $Z_V=Z_A$ which was also confirmed
numerically in Ref.~\cite{Bi:2017ybi}.

\subsubsection{$f_P$ and $f_V$}
To obtain decay constants $f_P$ and $f_V$ we perform simultaneous fits to the wall-to-point
($C_{P/V}(r=0, t)$) and wall-to-wall ($C^W(t)$) correlators for a given meson $M$.
These fittings are
with two exponentials and the ground state mass is constrained within $10\sigma$ to
its fitted result from the above strategy II as we determined the meson masses.
After removing the matrix element of the source operator $\langle0|O_\Gamma^{(W)}|M\rangle$
from the spectral weight of $C_{P/V}(r=0, t)$, we obtain $\langle0|O_\Gamma|M\rangle$
and then the decay constants $f_{P/V}$ by using Eqs.(\ref{eq:fp_pseudo},\ref{eq:fv})
and using the fitted meson mass.
This fitting and calculation process is repeated for each Jackknife sample to get the statistical uncertainty of $f_{P/V}$.
For $f_V$ obtained from the local vector current we need to multiply it with
$Z_V(=Z_A)$ as discussed in Section~\ref{sec:zavt}.
In the following we use a superscript ``bare" to indicate
decay constants obtained directly from the local vector current.

The bare decay constants $f_V$ in lattice units for the $\phi$ meson at our two strange quark masses
are given in the third column of Table~\ref{tab:kstar}. The two center values are almost the same and
our statistical uncertainty is big ($\sim6\%$). Thus, it is hard to tell
the strange quark mass dependence of $af_{\phi}^{\rm bare}$.
If we do a constant fit to the two numbers, then we obtain
$(af_{\phi}^{\rm bare})^{\rm phys}=0.1265(49)$ or $f_{\phi}^{\rm phys}=241(9)$ MeV
after multiplying it with $1/a$ and $Z_V$. If we do a linear extrapolation to
the physical strange quark mass point $a^2 m_{ss}^2(\rm phys)=0.1565$,
then we find a value $f_{\phi}^{\rm phys}=243.4(1.3)$ MeV. We choose the value with a larger error from the constant fit as our result at the physical point.
Therefore, we give
\begin{equation}
f_{\phi}^{\rm phys}=241(9)(2)\mbox{ MeV}
\label{eq:fphi}
\end{equation}
as our final result, where the second error comes from the difference between the constant fit and
the linear extrapolation and is treated as a systematic error.

We use $f_{D_s}$ to estimate our discretization error due to the large charm quark mass
since we cannot extrapolate to the continuum limit with only one lattice spacing.
The decay constant in lattice units $af_{D_s}$ for all charm and strange quark masses are
given in Table~\ref{tab:Ds}. One can use the function form given in Eq.~(\ref{eq:mDsstar})
(replacing $m_{D^*_s}$ with $f_{D_s}$)
to extrapolate/interpolate our lattice results in Table~\ref{tab:Ds}
to the physical charm and strange quark mass point.
What we find is
\begin{equation}
af_{D_s}^{\rm phys}=0.144(3)\quad\mbox{ or}\quad
f_{D_s}^{\rm phys}=249(5)\mbox{ MeV}.
\end{equation}
The difference in the center values of $f_{D_s}^{\rm phys}$
calculated in this work and in our previous work (254(2)(4) MeV)~\cite{Yang:2014sea} is 5 MeV or 2\%.
Since our previous result was obtained in the continuum limit,
we treat this 2\% difference as an estimate of the discretization error and assign it to
all our decay constants for the charmed mesons in this work.

\begin{figure}[htpb]
\centering\includegraphics[width=0.42\textwidth]{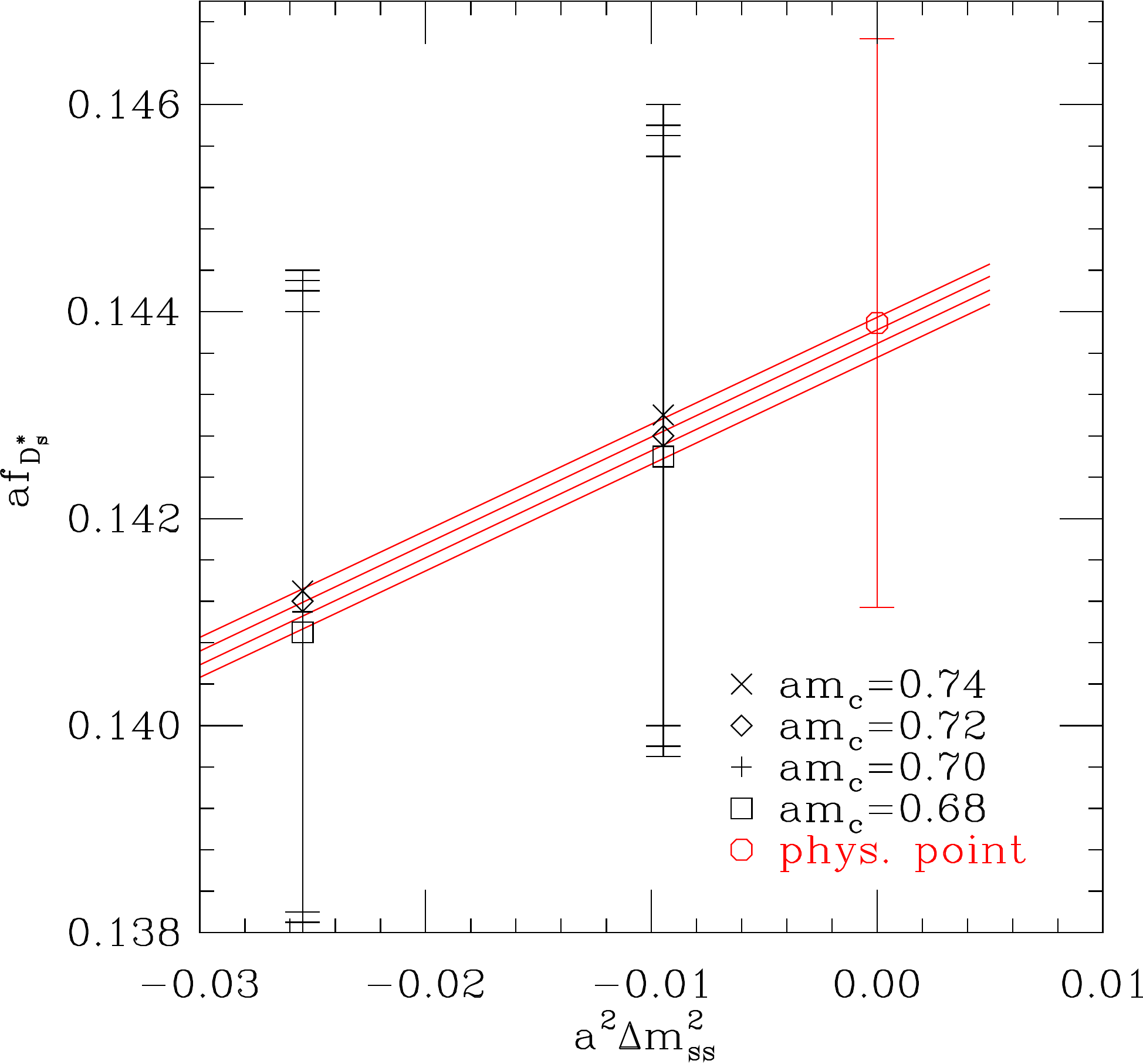}
\centering\includegraphics[width=0.43\textwidth]{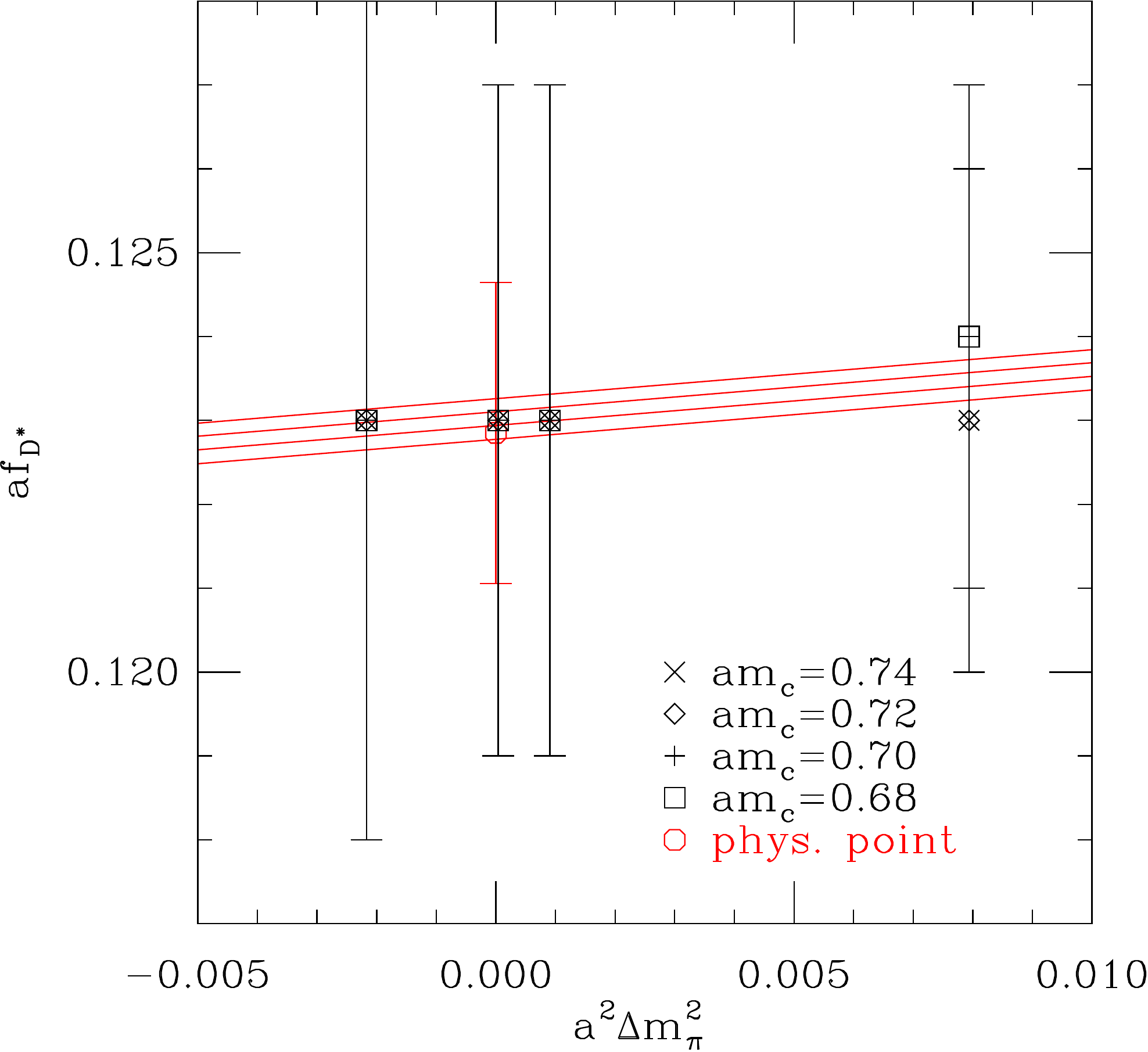}
\caption{The interpolation/extrapolation of $f_{D_s^*}$ to the physical point by using function form Eq.(\ref{eq:mDsstar}) (left panel). The right panel shows the interpolation/extrapolation of $f_{D^*}$ by using function form Eq.(\ref{eq:D}). The quark mass dependence is hard to see with the relatively big statistical errors.
The octagons show the results at the physical strange and charm quark mass point.}
\label{fig:fDstar_fDs_star}
\end{figure}

The vector meson decay constant $af_{D_s^*}^{\rm bare}$ from our lattice data is
given in the sixth column of Table~\ref{tab:Ds}.
Again we use the function form Eq.~(\ref{eq:mDsstar})
(replacing $m_{D^*_s}$ with $f_{D_s^*}$)
to extrapolate/interpolate our lattice results to the physical charm and strange
quark mass point. The fitting is shown on the left panel of Fig.~\ref{fig:fDstar_fDs_star}.
Compared with the case of $am_{D_s^*}$ the quark mass dependence of $af_{D_s^*}^{\rm bare}$
is hard to see with the relatively big statistical errors.

From the extrapolation/interpolation we get $(af_{D_s^*}^{\rm bare})^{\rm phys}=0.144(3)$.
Multiplying this number with $1/a=1.730(4)$ GeV and $Z_V=1.1025(16)$, we find
$f_{D_s^*}^{\rm phys}=274(5)$ MeV. Here the uncertainty includes the errors from
the statistics, extrapolation/interpolation, lattice spacing and $Z_V$.
If we assign a 2\% discretization error, then we finally get
\begin{equation}
f_{D_s^*}^{\rm phys}=274(5)(5)\mbox{ MeV}.
\label{eq:fDs_star}
\end{equation}

At each quark mass combination we find the ratio $f_{D_s^*}^{\rm bare}/f_{D_s}$ as given
in the last column of Table~\ref{tab:Ds}. The statistical error is from Jackknife
by using the Jackknife estimates of $f_{D_s^*}$ and $f_{D_s}$.
Then the ratio is extrapolated/interpolated to the physical quark mass point
by using the function form in Eq.(\ref{eq:mDsstar}) (replacing $m_{D^*_s}$ with the ratio). What we find is $(f_{D_s^*}^{\rm bare}/f_{D_s})^{\rm phys}=0.999(24)$.
Multiplying it with $Z_V$ and assigning a 2\% discretization error, we obtain
\begin{equation}
(f_{D_s^*}/f_{D_s})^{\rm phys}=1.101(27)(22).
\label{eq:rDs_hqs}
\end{equation}

The decay constants $f_{D}$ and $f_{D^*}$ and the ratio $f_{D^*}/f_{D}$ from our lattice data
are shown in Table~\ref{tab:D}. Similarly to the above analyses for $f_{D_s}$
and $f_{D_s^*}$, we get
\begin{equation}
f_{D}^{\rm phys}=213(2)(4)\mbox{ MeV},\quad\quad f_{D^*}^{\rm phys}=234(3)(5)\mbox{ MeV},
\label{eq:fDDstar}
\end{equation}
\begin{equation}
(f_{D^*}/f_{D})^{\rm phys}=1.10(2)(2).
\label{eq:rD_hqs}
\end{equation}
Here the first error comes from statistics and the interpolation/extrapolation to the physical quark mass point by using
Eq.(\ref{eq:D}) with the replacement of $m_{D^{(*)}}$ by the decay constants or their ratio.
For $f_{D^*}$ the error of $Z_V$ is also included in the first error.
The second error is the 2\%
systematic uncertainty due to the finite lattice spacing.
As an example, the interpolation of $f_{D^*}$ to the physical pion mass is shown in the right panel of Fig.~\ref{fig:fDstar_fDs_star}. Since our four light quark masses are distributed around and close to the physical point (the same is also true for our charm quark masses), the uncertainty of $f_{D^*}$ at the physical point is smaller than those of the lattice data.

Now we turn to the ratios $f_{D_s}/f_D$ and $f_{D_s^*}/f_{D^*}$ which reflect the size
of SU(3) flavor symmetry breaking. These ratios can be calculated in two ways.
One is using our final results for $f_{D_{(s)}^{(*)}}$ at the physical quark mass point.
By doing this we get $1.17(4)$ for both ratios. The other way is first calculating these ratios at
our nonphysical quark masses and then interpolating/extrapolating them to the physical
point by using the function form in Eq.(\ref{eq:D}). The second way gives
$f_{D_s}/f_D=1.163(14)$ and $f_{D_s^*}/f_{D^*}=1.17(2)$ without including the 2\%
discretization error. Including this error leads to
\begin{equation}
f_{D_s}/f_D=1.163(14)(23)\quad\mbox{and}\quad f_{D_s^*}/f_{D^*}=1.17(2)(2),
\label{eq:r_su3}
\end{equation}
which we take as our final results for the two ratios.
They tell us that SU(3) flavor symmetry breaking effects are of size $\sim17\%$.
Our value for $f_{D_s}/f_D$ agrees with the result from the RBC-UKQCD Collaborations in Ref.~\cite{Boyle:2018knm},
which uses unitary lattice setups with eight gauge ensembles including the 48I used in this work.

\subsubsection{$f_V^T/f_V$}
\begin{figure}[htpb]
\centering\includegraphics[width=0.48\textwidth]{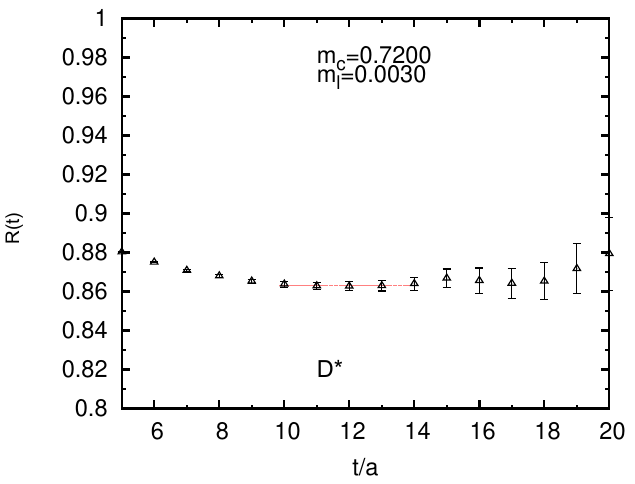}
\centering\includegraphics[width=0.48\textwidth]{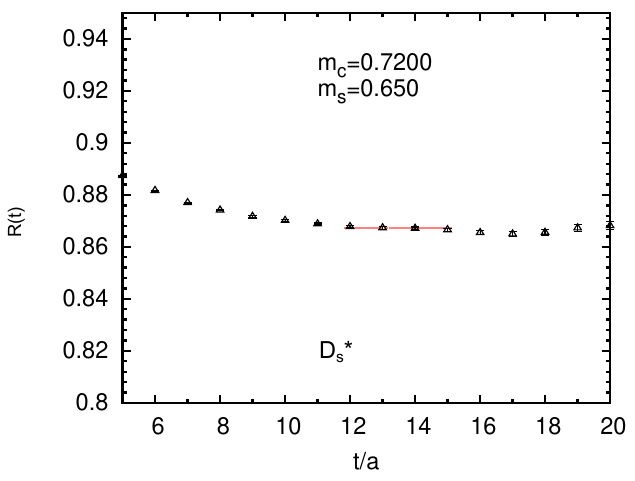}
\caption{Ratio of two-point functions $R(t)$ for $D^*$ (left panel) and $D_s^*$ (right panel).}
\label{fig:Rt}
\end{figure}

Because of the bad signal-to-noise ratio in $C_T(r=0, t)$
we do not directly determine the decay constant $f_V^T$ but calculate the ratio
$f_V^T/f_V$ from the ratio of two-point functions
\begin{equation}
\frac{f_V^T}{f_V}=\lim_{t\rightarrow \infty}\frac{C_T(r=0,t)}{C_V(r=0,t)}\equiv
\lim_{t\rightarrow \infty}R(t).
\end{equation}
The cancelation of statistical fluctuations from the numerator $C_T(r=0,t)$ and the denominator $C_V(r=0,t)$ leads to a better signal for the ratio $R(t)$ since both two-point functions are calculated on the same gauge ensemble and thus are correlated. At the large time limit the contributions from the higher states to the two-point functions are suppressed by their heavier masses. Then from Eq.(\ref{eq:fv}) and Eq.(\ref{eq:fvt}) one can derive that the ratio approaches $f_V^T/f_V$ for the ground state since the other factors in the numerator and the denominator cancel out.
Fig.~\ref{fig:Rt} shows the ratio $R(t)$ for $D^*$ and $D_s^*$ in the left and right
panel respectively. The uncertainties $\delta R(t)$ of the ratio shown in the figure are from Jackknife analyses.
As we can see, this ratio approaches a plateau at large $t$. We do constant fits
to $R(t)$ in the range [$t_{\rm min}$, $t_{\rm max}$] to get $f_V^T/f_V$,
where $t_{\rm max}$ is
fixed to the maximum value of $t$ with $\delta R/R\le10\%$.
$t_{\rm min}$ is varied to check the stability of the fitting results.
The variation ranges of $t_{\rm min}$ are indicated by the red lines in
Fig.~\ref{fig:Rt}. We make sure all the fittings give consistent results.
In this way we get the bare value of $f_V^T/f_V$ at each quark mass point. As an example,
the numerical results of this ratio for $D_s^*$ are presented in Table~\ref{tab:fvt_fv_fDsStar}.
\begin{table}
\begin{center}
\caption{Bare values of $f_{D_s^*}^T/f_{D_s^*}$ at various valence quark masses.}
\begin{tabular}{ccccc}
\hline\hline
  & \multicolumn{4}{c}{$am_c$}  \\
  & 0.6800 & 0.7000 & 0.7200 & 0.7400  \\
  \hline
   $am_s$ &  \multicolumn{4}{c}{$(f_{D_s^*}^T/f_{D_s^*})^{\rm bare}$} \\
  0.0580 & 0.862(2) & 0.865(2) & 0.867(2) & 0.869(2) \\
  0.0650 & 0.863(2) & 0.865(2) & 0.867(2) & 0.869(2) \\
 \hline\hline
\end{tabular}
\label{tab:fvt_fv_fDsStar}
\end{center}
\end{table}

Then we use Eq.(\ref{eq:mDsstar}) and Eq.(\ref{eq:D}) to interpolate/extrapolate
our raw data to the physical quark mass point for $f_{D_s^*}^T/f_{D_s^*}$
and $f_{D^*}^T/f_{D^*}$ respectively. After multiplying the results with
the renormalization factor $Z_T/Z_A(2\mbox{ GeV})=1.055(31)$ in the $\msbar$ scheme
and assigning a 2\% discretization uncertainty, we find
\begin{equation}
(f_{D_s^*}^T/f_{D_s^*})^{\rm phys}=0.92(3)(2)\quad \mbox{and}\quad
(f_{D^*}^T/f_{D^*})^{\rm phys}=0.91(3)(2)
\label{eq:r_fT}
\end{equation}
at the scale 2 GeV.
Here the first uncertainty includes the errors from statistics and
interpolation/extrapolation and the error of $Z_T/Z_A(2\mbox{ GeV})$,
and is dominated by the error of the renormalization factor.
The second uncertainty is from the finite lattice spacing effect.

\section{Summary}
\label{sec:summary}
We calculated the decay constants $f_P$, $f_V$ and $f_V^T/f_V$ of the charmed
and light mesons including $D_{(s)}^{(*)}$ and $\phi$
by using
2+1-flavor domain wall fermion gauge configurations at one lattice spacing.
The valence overlap fermion has 4, 2 and 4 mass values respectively for
the light, strange and charm quarks.
We use the experiment values of $m_\pi$, $m_{ss}^2\equiv 2m_K^2-m_\pi^2$ and $m_{D_s}$ to
set the physical light, strange and charm quark masses.
The masses of $D$, $D_{(s)}^*$, $\phi$ and $K^*$ at the physical point are found by interpolation/extrapolation
using the lowest order of Taylor expansion (i.e., a linear interpolation/extrapolation)
since our valence quark masses are close to their physical values.

The masses $m_D$, $m_{D_s^*}$, $m_\phi$ and $m_{K^*}$ obtained from our lattice calculation are in good
agreement with their experiment measurements.
The $D^*$ mass we found is 1\% higher than its experiment value.
The center value of $f_{D_s}$ from this calculation is 2\% away from our previous lattice QCD calculation
extrapolated to the continuum limit~\cite{Yang:2014sea}.
Thus, we estimate the discretization uncertainty in this work to be around 2\%.

The final results of this work for the decay constants are given in
Eqs.(\ref{eq:fphi},\ref{eq:fDs_star}-\ref{eq:r_su3},\ref{eq:r_fT}).
Quadratically adding together the statistical/fitting uncertainty and the systematic uncertainty,
we get the decay constants in Table~\ref{tab:final1}
and some of their ratios in Table~\ref{tab:final2}.
\begin{table}
\begin{center}
\caption{Decay constants of $D_{(s)}^{(*)}$ and $\phi$
in units of MeV. $f_V^T/f_V$ is given in the $\msbar$ scheme
at the scale 2 GeV.}
\begin{tabular}{cccccc}\hline\hline
 & $D_s$ & $D_s^*$ &  $D$ & $D^*$ & $\phi$ \\
\hline
$f_{P/V}$/MeV & 249(7)  & 274(7) & 213(5)  & 234(6) & 241(9) \\
$f_V^T/f_V$ & -  & 0.92(4)  & - & 0.91(4) & - \\
  \hline
  \hline
\end{tabular}
\label{tab:final1}
\end{center}
\end{table}
For the light vector meson $\phi$ the statistical error dominates the uncertainties.
While for the heavy mesons the discretization error and the error from $Z_T/Z_A$ (when needed)
are the main sources of uncertainty.
We believe our results for $f_{D_s^*}^T/f_{D_s^*}$ and $f_{D^*}^T/f_{D^*}$ are the first lattice QCD calculations, which can be used as input parameters for LCSR calculations of
form factors in $B$ to vector meson semileptonic decays.
\begin{table}
\begin{center}
\caption{Ratios of decay constants for $D_{(s)}^{(*)}$.}
\begin{tabular}{cccc}\hline\hline
$f_{D^*}/f_D$ & $f_{D_s^*}/f_{D_s}$ & $f_{D_s}/f_{D}$ &  $f_{D_s^*}/f_{D^*}$  \\
\hline
1.10(3) & 1.10(4)  &  1.16(3) & 1.17(3)    \\
  \hline
  \hline
\end{tabular}
\label{tab:final2}
\end{center}
\end{table}

Our number $f_\phi=241(9)$ MeV is lower than the $N_f=2$ lattice simulation result
in Ref.~\cite{Jansen:2009hr}, which gives $f_\phi=308(29)$ MeV. This may be due to the dynamical
strange quark effects. Note our $f_\phi$ is in good agreement with that in~\cite{Donald:2013pea}, which is also a
2+1-flavor lattice calculation. The experimental value of $f_\phi$ can be extracted from
$\Gamma(\phi\rightarrow e^+ e^-)=1.251(21)$ keV~\cite{Tanabashi:2018oca} by using the relation
\begin{equation}
\Gamma(\phi\rightarrow e^+ e^-)=\frac{4\pi\alpha^2_{\rm em}}{27m_\phi} f_\phi^2.
\end{equation}
Inputting $\alpha_{\rm em}=1/137.036$ and $m_\phi=1019.461(16)$ MeV~\cite{Tanabashi:2018oca}, one finds $f_\phi^{\rm exp}=227(2)$ MeV. Our result agrees with the experiment value at $1.5\sigma$.

Our value for $f_D$ is 213(5) MeV, which agrees with other lattice QCD calculations with 2-flavor~\cite{Carrasco:2013zta},
2+1-flavor~\cite{Bazavov:2011aa,Na:2012iu,Boyle:2017jwu}
and 2+1+1-flavor~\cite{Bazavov:2017lyh,Carrasco:2014poa} simulations.
Combining the latest experimental average $f_{D^+}|V_{cd}|=45.91(1.05)$ MeV from PDG2018~\cite{Tanabashi:2018oca} and our $f_D=213(5)$ MeV, one gets
\begin{equation}
|V_{cd}|=0.2155(51)(49).
\end{equation}
Here the two errors are from the lattice calculation and experiment, respectively.

\begin{figure}[htpb]
\centering\includegraphics[width=0.421\textwidth]{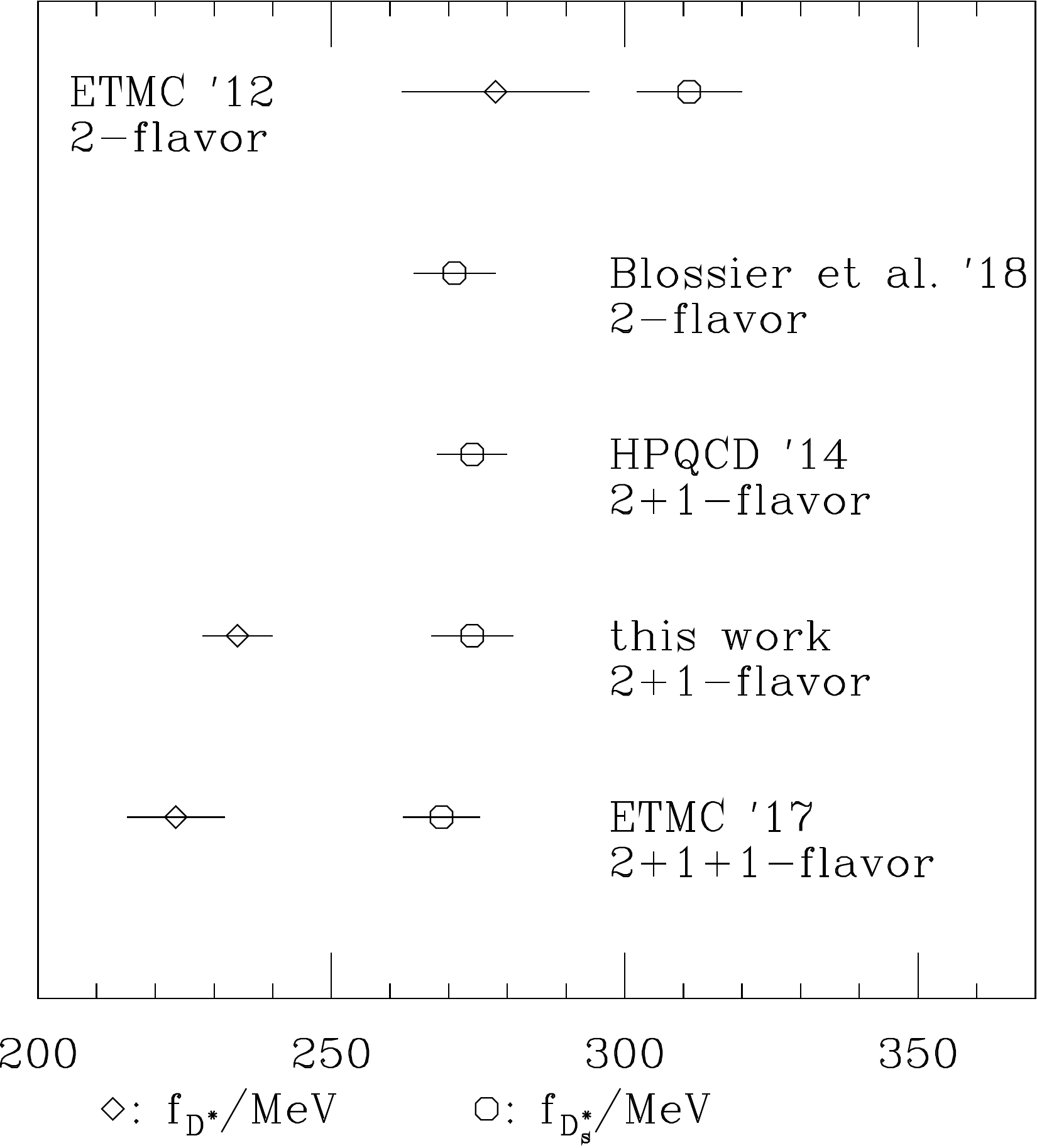}
\centering\includegraphics[width=0.43\textwidth]{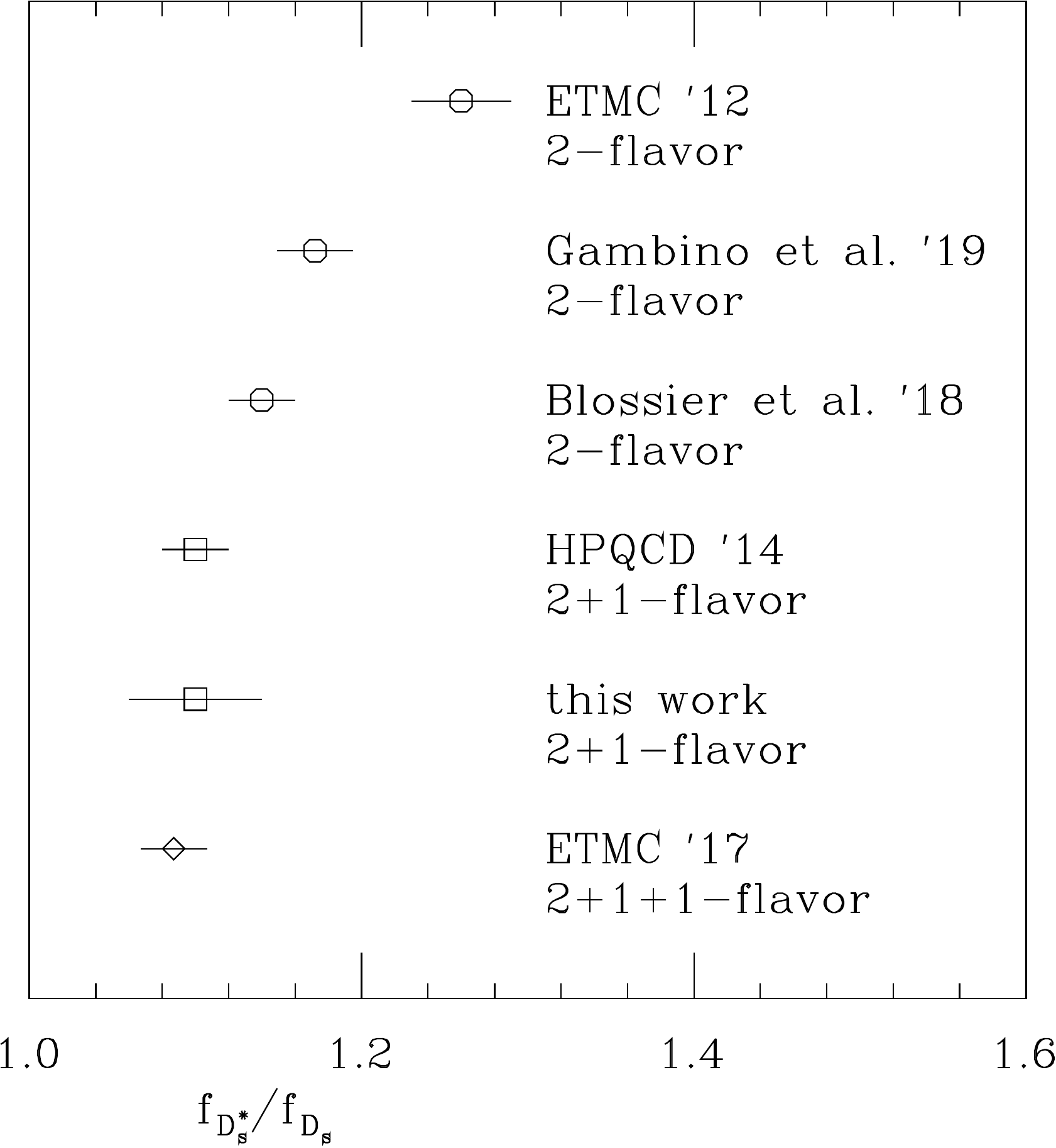}
\caption{Comparisons of $f_{D_{(s)}^*}$ (left panel) and $f_{D_s^*}/f_{D_s}$ (right panel)
from lattice QCD calculations.}
\label{fig:comparison}
\end{figure}

In Fig.~\ref{fig:comparison} we compare $f_{D_{(s)}^*}$ and the ratio
$f_{D_s^*}/f_{D_s}$ from this work and other lattice QCD calculations~\cite{Donald:2013sra,Becirevic:2012ti,Blossier:2018jol,Lubicz:2017asp,Gambino:2019vuo}.
The values from 2+1-flavor and 2+1+1-flavor simulations are in consistency.
There might be a
tension between 2-flavor calculations and the other calculations including the dynamical strange quark. This may reflect an unexpected large quenching effect from the strange quark. However the 2-flavor calculation of $f_{D_s^*}/f_{D_s}$ in~\cite{Blossier:2018jol} shows that this quenching effect is not so significant as that seen in~\cite{Becirevic:2012ti}. The two calculations employ different lattice actions of the two-flavor theory. The computation in~\cite{Gambino:2019vuo} is performed on the same 2-flavor gauge ensembles as used in~\cite{Becirevic:2012ti} and employs the analysis method as used in~\cite{Lubicz:2017asp}. It gives a $f_{D_s^*}/f_{D_s}$ with a smaller strange quark quenching effect, and therefore is more in agreement with~\cite{Blossier:2018jol}.
Thus, more lattice QCD calculations, especially those with two dynamical flavors, are certainly welcome to clarify this situation.

The ratios of decay constants of charmed mesons in Table~\ref{tab:final2} show that the size of heavy quark symmetry breaking is about $10\%$. While the size of SU(3) flavor symmetry breaking is around $17\%$.

To better control the systematic uncertainty from discretization effects in our work, we need to perform
our calculation at more lattice spacings in the future. Also we need to include the quark-line
disconnected diagram for the $\phi$ meson two-point function. To accurately estimate the threshold effects of
strong decays of vector mesons, further studies on larger volumes are necessary.

\section*{Acknowledgements}
This work was supported by the National Key Research and Development Program of China (No. 2017YFB0203200).
We thank RBC-UKQCD collaborations for sharing the domain wall fermion configurations.
This work was partially supported
by the National Natural Science Foundation of China (NSFC) under Grant 11935017.
This research used resources of the Oak Ridge Leadership Computing
Facility at the Oak Ridge National Laboratory, which is supported by the
Office of Science of the U.S. Department of Energy under Contract No.
DE-AC05-00OR22725. This work used Stampede time under the Extreme
Science and Engineering Discovery Environment (XSEDE), which is
supported by National Science Foundation Grant No. ACI-1053575.
We also thank the National Energy Research Scientific Computing Center (NERSC) for providing HPC
resources that have contributed to the research results reported within this paper.

\end{document}